\documentclass[%
 reprint,
 amsmath,amssymb,
 aps,
floatfix,showkeys,
]{revtex4-2}

\usepackage{graphicx}
\usepackage{dcolumn}
\usepackage{bm}
\usepackage{color}
\usepackage{wrapfig}
\usepackage{physics}
\usepackage{fancyhdr}
\usepackage{lastpage}
\usepackage{amsmath}

\newcommand{\be}{\begin{equation}}
\newcommand{\ee}[1]{\label{#1} \end{equation}}

\def\_#1{\textsubscript{#1}}
\def\^#1{\textsuperscript{#1}}

\usepackage[symbol]{footmisc}

\begin{document}

\preprint{APS/123-QED}

\title{Thermal runaway of silicon-based laser sails}

\author{Gregory R. Holdman$^{*}$ and Gabriel R. Jaffe}
\thanks{The authors contribute equally to this paper.}
\affiliation{Department of Physics, University of Wisconsin-Madison, Madison WI 53706 USA}

\author{Min Seok Jang}

\affiliation{
School of Electrical Engineering, Korea Advanced Institute of Science and Technology, Daejeon 34141, Korea
}%

\author{Demeng Feng and Mikhail A. Kats}

\affiliation{
Department of Electrical and Computer Engineering, University of Wisconsin-Madison, Madison WI 53706 USA
}%

\author{Victor Watson Brar}
\thanks{vbrar@wisc.edu}%
\affiliation{%
Department of Physics, University of Wisconsin-Madison, Madison WI 53706 USA
}%

\date{\today}

\begin{abstract}
\begin{center}
\textbf{\abstractname}

\end{center}

Laser sail-based spacecraft -- where a powerful earth-based laser propels a lightweight outer-space vehicle -- have been recently proposed by the Breakthrough Starshot Initiative as a means of reaching relativistic speeds for interstellar spacetravel.  The laser intensity at the sail required for this task is at least 1 GW$\cdot$m$^{-2}$ and, at such high intensities, thermal management of the sail becomes a significant challenge even when using materials with low absorption coefficients.  Silicon has been proposed as one leading candidate material for the sail due to its low sub-bandgap absorption and high index of refraction, which allows for low-mass-density designs.  However, here we show that the temperature-dependent bandgap of silicon combined with two-photon absorption processes can lead to thermal runaway for even the most optimistic viable assumptions of the material quality.  From our calculations, we set bounds on the maximum laser intensities that can be used for a thermally stable, Si-based laser sail.

\end{abstract}

\maketitle

Sending probes to nearby star systems requires engineering spacecraft that can travel at relativistic speeds.  Recently, the Breakthrough Starshot Initiative\cite{BreakthroughInitiatives} has proposed to design, construct, and launch `laser sails' in order to achieve this goal, with an aim of sending a spacecraft towards Alpha Centauri at 20\% the speed of light.  The proposed propulsion mechanism for these vehicles is the radiation pressure from a ground-based laser array on a reflective `sail'. Many design challenges must be overcome before such a project can be realized, and amongst the forefront of them is designing the sail, which must have high reflectivity, low mass, low absorption, lateral and rotational stability, and thermal stability. Optical metasurfaces and photonic crystals are ideally suited structures for laser sails because they can be made highly reflective \cite{Slovick_Yu_Berding_Krishnamurthy_2013}, have low areal densities \cite{Atwater_Davoyan_Ilic_Jariwala_Sherrott_Went_Whitney_Wong_2018}, and can achieve self-correcting stability when accelerated by a high-powered optical beam\cite{Siegel_Wang_Menabde_Kats_Jang_Brar_2019,Ilic_Atwater_2019}. However, absorption of even a tiny fraction of the incoming laser radiation, which will be in the range of 1-10\,GW$\cdot$m$^{-2}$, poses major issues for sail thermal management because thermal radiation is the only mechanism for cooling in space. To prevent the sail from melting, it is crucial to choose materials that balance exceptionally low absorption coefficients at the drive laser wavelength (including the Doppler shift after the sail has accelerated) in the near infrared and high absorption coefficients in the midinfrared to far infrared to increase radiative cooling \cite{Atwater_Davoyan_Ilic_Jariwala_Sherrott_Went_Whitney_Wong_2018, ilic_went_atwater_2018}.

Silicon and silicon dioxide have emerged as two of the most attractive materials for laser sail designs based on metasurfaces, owing to the large contrast in their refractive indices, low material absorption at the proposed driving laser wavelengths ($\lambda_0>1.2\,\mu$m)\cite{Atwater_Davoyan_Ilic_Jariwala_Sherrott_Went_Whitney_Wong_2018}, low mass densities, and potential for fast integration with existing industrial silicon fabrication infrastructure \cite{ilic_went_atwater_2018,Salary_Mosallaei_2020}. In particular, one study of optimized sail structures found that a one-dimensional Si grating provided superior acceleration performance over SiN$_x$\cite{Jin_Li_Orenstein_Fan_2020}. Modeling of laser heating of Si/SiO$_2$ sails with ~25\,GW$\cdot$m$^{-2}$ incident laser intensity have predicted equilibrium temperatures below the melting points of the constituent materials assuming that the absorption coefficient of both materials is $<$\,10$^{-2}$\,cm$^{-1}$\cite{ilic_went_atwater_2018}. However, those thermal calculations did not take into account two important factors that can significantly affect the thermal stability of the sail.  First, silicon exhibits a strong temperature-dependent absorption coefficient due to bandgap narrowing that, at a temperature of 800\,K and wavelength of 1.55\,$\mu$m, can reach $>$4.5\,cm$^{-1}$ \cite{Rogne_APL_1996}.  This value is $\sim$6 orders of magnitude higher than at room temperature\cite{Degallaix_OL_2013}. Second, two-photon absorption (TPA) introduces additional absorption in silicon with a TPA coefficient of $\beta=$1.35\,cm$\cdot$GW$^{-1}$ at 1550 nm, which would cause the absorption of the sail to increase with laser power even at low temperatures\cite{Bristow_APL_2007}. For laser-phased arrays operating at intensities on the order of $I_0\approx10$ GW$\cdot$m$^{-2}$, TPA adds about 1.35$\times 10^{-3}$\,cm$^{-1}$ to the absorption coefficient of Si, well above any absorption measured in Si at room temperature for low laser intensities. In this work, we show that these two properties can cause thermal instabilities in the sail during its acceleration phase.

To investigate the limits of thermal stability, we analyzed a Si/SiO$_2$-based metasurface laser sail that is designed to have a reflection coefficient of $>$99\% at the laser wavelength of 1.55\,$\mu$m and that satisfies the weight constraint of  $<$1\,g/m$^2$\cite{BreakthroughInitiatives}.  In order to model the thermal characteristics of the sail, we assembled detailed models of the spectral and temperature-dependent absorptivity of both Si and SiO$_2$ from literature sources and used these in our simulations.  We then used full-wave, finite element simulations to calculate the precise temperature dependence of the emissivity and absorbtivity, and we balanced these two factors to determine the stable sail temperature. We found that, even when using the lowest absorption coefficients ever demonstrated on the wafer scale, there is no equilibrium temperature that exists for laser intensities above 5.4 GW$\cdot$m$^{-2}$. The mechanisms behind this are TPA in Si at low temperatures and free-carrier absorption augmented by a decreasing bandgap at high temperatures also in Si. For laser intensities $<$5.4 GW$\cdot$m$^{-2}$ we found that the sail can exhibit a stable equilibrium temperature, but fluctuations above a higher unstable equilibrium temperature could lead to thermal runaway.

\begin{figure}[t]
\includegraphics[width = .9\linewidth]{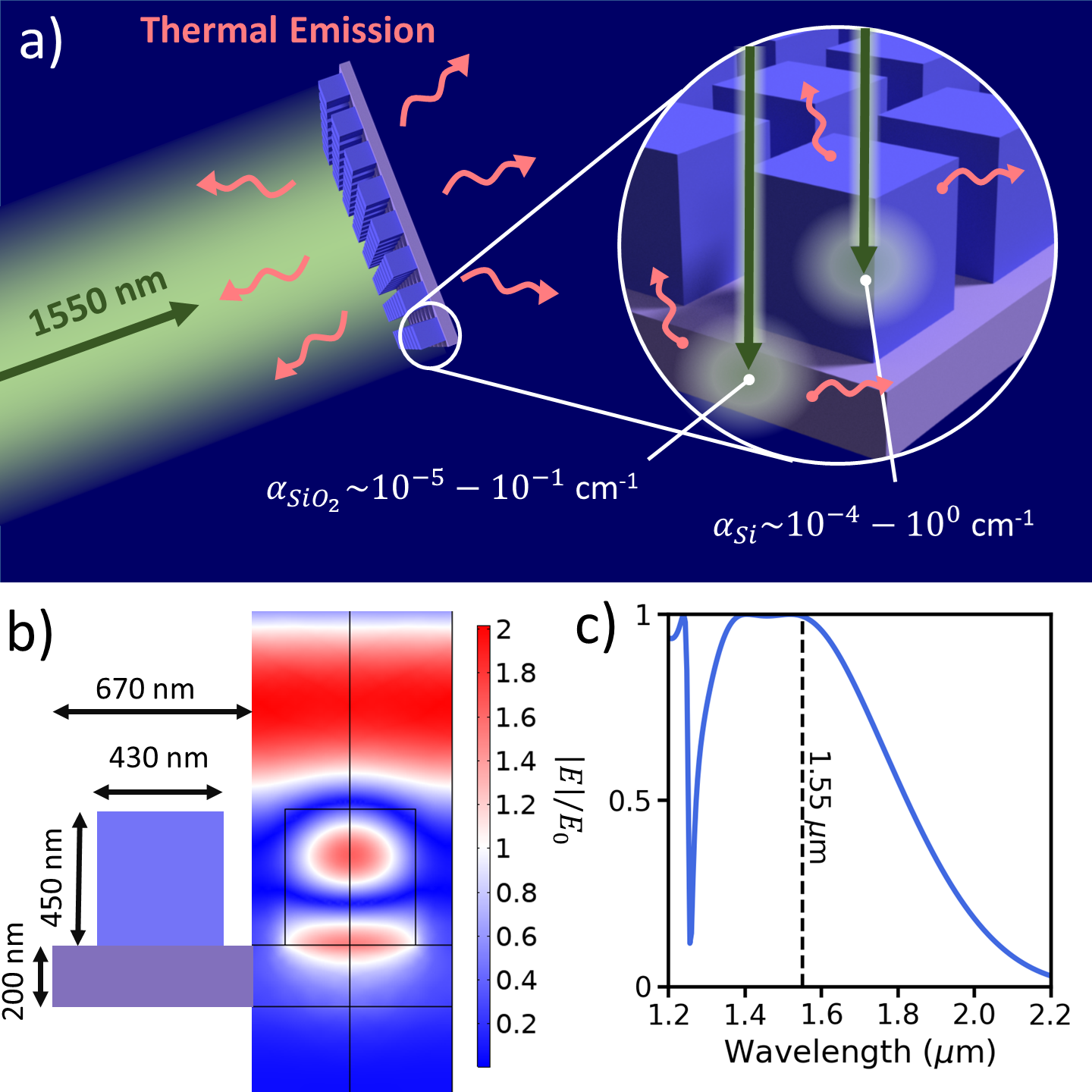}
\caption{a) Laser-sail probe being accelerated toward Alpha Centauri. Laser light is absorbed in the structure, which can only cool via thermal emission. Absorption coefficients at 1.55\,$\mu$m are labeled. There is wide variability in these materials for different qualities, temperatures, and intensities. b) Diagram of the metasurface used in this work consisting of 430-nm-wide by 450-nm-tall Si blocks (dark blue) with a pitch of 670 nm on a 200-nm-thick SiO$_2$ substrate (light purple). The electric field magnitude at 1550 nm is shown. c) Reflection spectrum of the metasurface.
\vspace{-2mm}}
\label{fig:absem}
\end{figure}

\vspace{-10pt}
\section*{\label{sec:matmodels}Sail and Material Models} 
\vspace{-10pt}

A schematic of the metasurface laser sail design used in this work is shown in Fig. \ref{fig:absem}.  Our Si/SiO\_{2}-based metasurface consisted of 430-nm-wide by 450-nm-tall Si blocks on a 200-nm-thick SiO$_2$ substrate, with a total areal mass density of 0.96 g/m$^2$.  This structure is based on a previously published broadband perfect reflector geometry that uses Mie resonances to achieve perfect reflection \cite{Slovick_Yu_Berding_Krishnamurthy_2013}.  In this work, we have re-optimized that structure to account for a finite-thickness silica layer, and achieved a reflectivity of 99.5 \% at a driving laser wavelength $\lambda_0=$1550nm and $>$95 \% from 1350\,nm to 1605\,nm.  We note that the thermal balance calculations we present here do not strongly depend on the structure geometry, and that similar results are obtained for continuous, unpatterned Si/SiO\_2 (see Supplement).

\begin{figure}[t]
\includegraphics[width =1\linewidth]{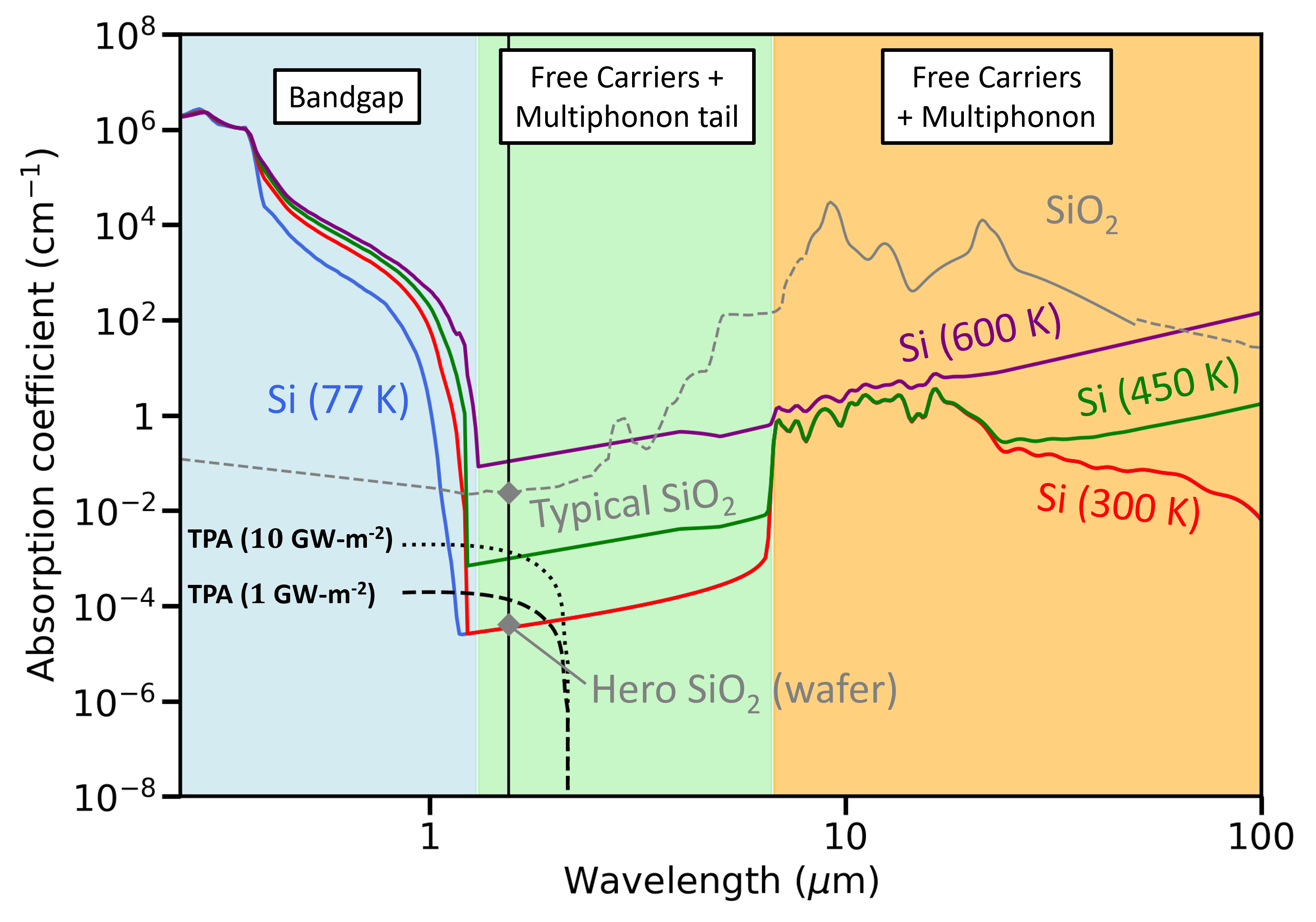}
\caption{A composite model from multiple literature sources of the absorption coefficient of Si as a function of wavelength at various temperatures (colored lines) and amorphous SiO$_2$ at room temperature (grey line). The blue, green, and orange shaded regions indicate the dominant absorption mechanisms for each wavelength range for Si.  The model for absorption in the bandgap region is from Ref.\ \cite{Green_SEMSC_2008}, the free-carrier models are from \cite{Rogne_APL_1996} and \cite{Schroder_IEEEJSSC_1978}, the multiphonon-absorption bands are from \cite{Johnson_1959,Wollack_Cataldo_Miller_Quijada_2020}. Two-photon absorption (TPA) at intensities 1 GW$\cdot$m$^{-2}$ and 10 GW$\cdot$m$^{-2}$ are included as dashed black lines \cite{Bristow_APL_2007}.  The SiO$_2$ absorption values are from Ref.\ \cite{kitamura_pilon_jonasz_2007}. The data point labeled "Hero" indicates the lowest demonstrated absorption value of SiO$_2$ at a wavelength of 1.55\,$\mu$m in a wafer \cite{Lee_Chen_Li_Yang_Jeon_Painter_Vahala_2012}. The vertical line indicates the laser wavelength of 1.55\,$\mu$m. Further details regarding the Si composite absorption model are available in the supporting information.}
\label{fig:matmodels}
\end{figure}

Our simulations included models of the complex refractive indices of both materials.  We assumed a constant real refractive index of $\text{Re}[n_{Si}]$=3.42 for Si as the index varies very little over the wavelength range of interest 1-100 $\mu$m. For the real refractive index of SiO$_2$ $\text{Re}[n_{SiO_2}]$, we followed the model in Ref. \cite{kitamura_pilon_jonasz_2007}. The imaginary parts of the refractive indices require more care. In order to calculate the equilibrium temperature of a laser-heated Si/SiO$_2$ metasurface sail, comprehensive models of the temperature and wavelength dependence of the Si and SiO$_2$ absorption coefficients $\alpha=4\pi\text{Im}[n]/\lambda$ are needed.  These models must be valid in the range of the expected equilibrium temperatures spanning 50--800\,K, and cover both the Doppler-broadened laser wavelength range of 1.55--1.90\,$\mu$m and the bandwidth of the thermal emission from 2--100\,$\mu$m.  We found no single literature source that provided absorption values over such a broad range of wavelengths and temperatures.  We therefore assembled a composite absorption model for each material from multiple literature sources, as seen in Fig\,\ref{fig:matmodels}.

There are four different processes that contribute to the infrared absorption of Si. The total absorption, $\alpha_{Tot}(\lambda,T,I_0)$, at a particular wavelength $\lambda$, temperature $T$, and incident intensity $I_0$, can be modeled as the sum of the absorption contributions from each absorption mechanism as
\begin{equation}
\begin{aligned}
\alpha_{Tot}(\lambda,T,I_0) = \alpha_{BG}(&\lambda,T)\ +\ \alpha_{FC}(\lambda,T)\ +\\
&\alpha_{TPA}(\lambda,I_0)\ +\ \alpha_{L}(\lambda).
\end{aligned}
\end{equation}
Here, $\alpha_{BG}$ is the coefficient of the bandgap absorption caused by the excitation of electrons in the valence band into the conduction band, and $\alpha_{FC}$ is the coefficient of the free-carrier absorption, which involves the transfer of photon energy to thermally excited free carriers in conduction or valence bands. The two-photon absorption coefficient is $\alpha_{TPA}$, and $\alpha_{L}$ is the coefficient for the lattice absorption through multiphonon processes.

\begin{figure*}
\centering

\includegraphics[width=.8\textwidth]{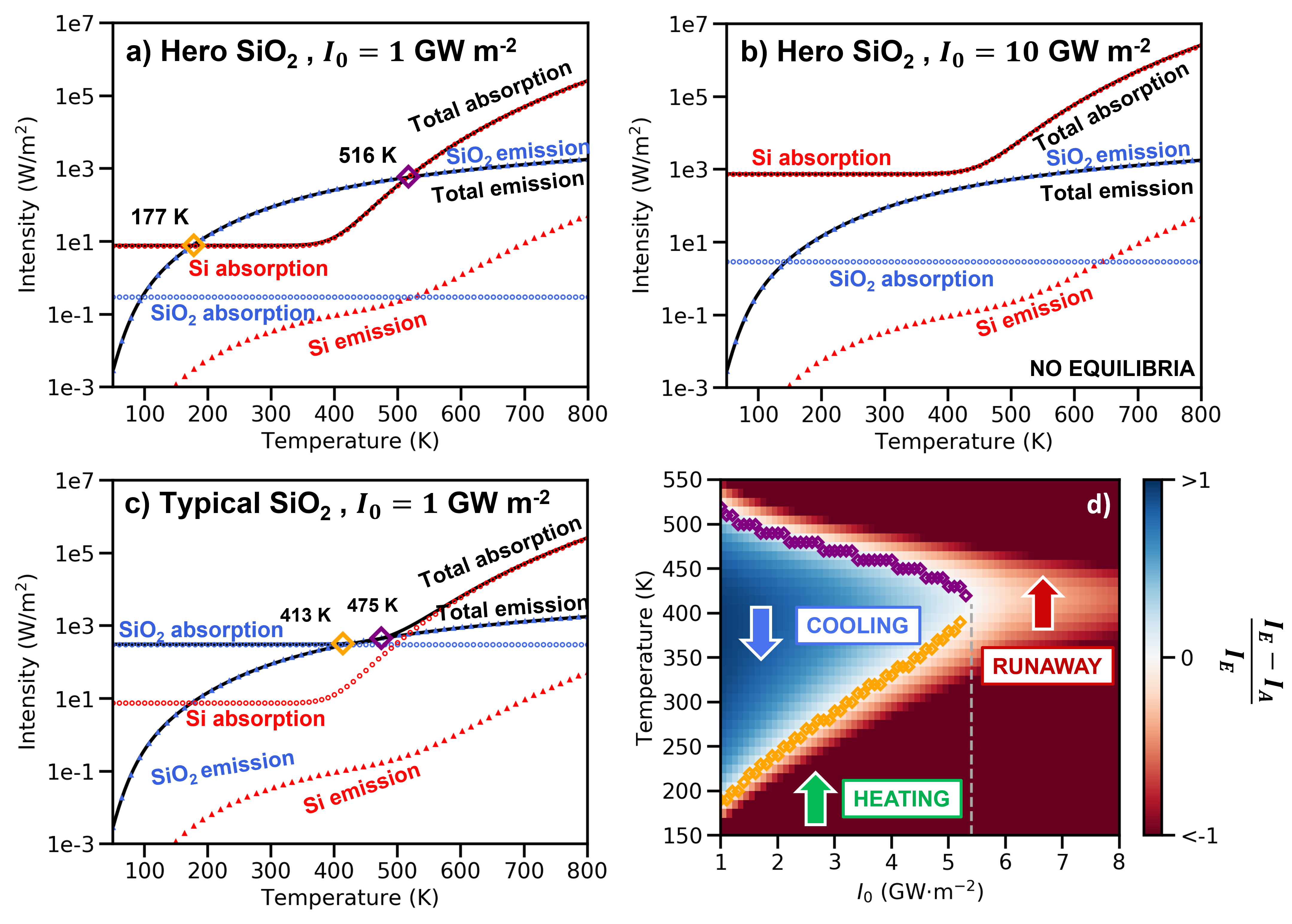}

\caption{(a-c) Absorbed and thermally emitted intensities versus temperature for the Si/SiO$_2$ metasurface depicted in Fig.\,\ref{fig:absem} made from either Hero or Typical SiO$_2$ and illuminated with an incident laser intensity I$_0$. Stable (orange) and unstable (purple) equilibrium temperatures are labeled as diamonds.  Note, there is no equilibrium temperature for Hero SiO$_2$ at I$_0 = $10\,GW$\cdot$m$^{-2}$.  Absorbed and emitted intensities for the complete laser sail are shown in black while the contribution of the Si and SiO$_2$ layers are shown in red and blue, respectively. Dotted lines are used to indicate the Si absorption and SiO$_2$ emission because they overlap with the total absorption and emission. (d) Normalized difference in emissive intensity and absorptive intensity as a function of temperature $T$ and incident intensity $I_0$ for a sail made with Hero SiO$_2$. Stable and unstable equilibrium temperatures are indicated as orange and purple diamonds, respectively. The blue and green arrows indicate the zones where the sail will either cool or heat to a stable equilibrium temperature below $I_0\approx5.4$\,GW$\cdot$m$^{-2}$. Otherwise, thermal runaway occurs.}

\label{fig:intensities}
\end{figure*}

The composite absorption model for Si that we developed includes tabulated values of the total absorption coefficient between 0.25--1.45\,$\mu$m\cite{Green_SEMSC_2008}, a free-carrier absorption model spanning 1.45--5\,$\mu$m\cite{Rogne_APL_1996}, and a second free-carrier model for the range of 4--100\,$\mu$m \cite{Schroder_IEEEJSSC_1978} for which we assume the intrinsic-carrier concentration follows the model in Ref.\,\cite{Couderc_JAP_2014}.  We also include the multiphonon-absorption bands from 7--100\,$\mu$m \cite{Johnson_1959,Wollack_Cataldo_Miller_Quijada_2020} as well as tabulated coefficients for TPA which has a broad step near 2\,$\mu$m \cite{Bristow_APL_2007}.The TPA process is a particularly important mechanism to consider when modeling metasurfaces designed to operate at high intensities.  In Si at 300\,K, the TPA at 1.55\,$\mu$m is greater than the combined free-carrier and single-photon bandgap absorption for fluences $>$900\,kW$\cdot$m$^{-2}$.  The calculation of $\alpha_{TPA}$ in this work assumes the incident beam intensity is uniform everywhere and therefore does not include possible field enhancement effects within the metasurface. Further information regarding interpolation between models and the fits to the multiphonon modes is available in the supporting information.

The amorphous SiO$_2$ absorption model we assembled includes the extensive collation of room-temperature wavelength-dependent absorption measurements of SiO$_2$ glass from 0.015--100\,$\mu$m found in Ref.\,\cite{kitamura_pilon_jonasz_2007}. The room-temperature values are used at all sail temperatures in our models because the absorption of SiO$_2$ exhibits only a weak temperature dependence from 300--1100\,K between 1--7\,$\mu$m \cite{Edwards_NASA_1966,Dvurechensky_IP_1979,Tan_JPCS_2000}.  We note that the growth conditions, defect density, and concentration of hydroxyl groups are known to have a strong effect on the absorption of SiO$_2$ in the infrared and are responsible for the significant variation in the reported absorption of SiO$_2$ at 1.55\,$\mu$m. Films of SiO$_2$ grown via wet oxidation of Si followed by a 24 hour dry oxidation at 1000\,\textdegree C have demonstrated absorption as low as  $4.1\times$10$^{-5}$\,cm$^{-1}$ on the wafer scale\cite{Lee_Chen_Li_Yang_Jeon_Painter_Vahala_2012}. This absorption value is three orders of magnitude smaller than the typical value of $2.4\times$10$^{-2}$\,cm$^{-1}$\cite{kitamura_pilon_jonasz_2007}. Even lower values of absorption have been demonstrated in optical fibers, but such low values of absorption have not been realized at the wafer scale. We refer to the above absorption coefficients at 1.55\,$\mu$m as ``Hero" and ``Typical", respectively, and they are denoted in Fig.\ \ref{fig:matmodels} by grey diamonds.  To our knowledge, the ultra-low Hero SiO$_2$ absorption has only been demonstrated in a single experiment.

\vspace{-10pt}
\section*{\label{sec:sim}Simulation of Metasurface Thermal Balancing} 
\vspace{-10pt}

During the acceleration phase, the sail absorbs energy from the driving laser while the only mechanism for losing energy is thermal radiation. The absorbed laser power will raise the temperature of the sail until it reaches an equilibrium temperature $T_i$, at which point the absorbed intensity $I_A$ and thermally emitted intensity $I_E$ are equal.  Perturbations may drive the sail above or below $T_i$, but if a rising temperature causes $I_E(T)$ to exceed $I_A(T)$, the equilibrium temperature will be stable because excess emission cools the sail. Likewise if $I_A(T)$ increases above $I_E(T)$ with $T$, then $T_i$ is an unstable equilibrium as the sail absorbs more energy than it can emit and it undergoes thermal runaway. It is possible that $I_A(T)>I_E(T)$ for a large range, possibly even all values of $T$. In this case, the temperature will increase until the sail melts or is otherwise destroyed. In this work, we predicted the equilibrium temperatures of the sail in Fig.\,\ref{fig:absem}a by calculating $I_A(T)$ and $I_E(T)$ for a large range of temperatures and then identified the $T_i$ where $I_A(T_i)\approx I_E(T_i)$.

 We computed the quantities $I_A(T)$ and $I_E(T)$ in different ways. Absorptivity ($A(T,I_0)=I_A(T,I_0)/I_0$) at the driving laser wavelength 1.55\,$\mu$m was calculated using a full-wave simulation in COMSOL Multiphysics. We swept $T$ and $I_0$  to calculate $A(T,I_0)$ via a volumetric integration of absorptive loss. While a more accurate nonlinear simulation would recalculate the index at each coordinate in the Si region, we found such a computationally expensive approach to be unnecessary because the average electric field magnitude in the Si block is within 15\% of the incident field. For $I_E(T)$, it is computationally challenging to calculate the exact thermal emission of the metasurface structure using COMSOL given that light is emitted over a broad bandwidth ($\lambda\sim$2-100 $\mu$m), over all angles, and with both $s$ and $p$ polarizations. Thus, to calculate $I_E(T)$, we approximated the Si layer as a material with an effective index $n_{eff}=F\times n_{Si}$ where $F=430^2/670^2\approx 41\%$ is the fill factor of the Si in our metasurface design, and $n_{Si}$ is the bulk Si refractive index. This approach leverages the fact that the characteristic length scale of the metasurface is much smaller than the emission wavelengths. We confirmed the validity of this method by comparing full-wave simulations of the metasurface to the approximated structure for a few chosen angles (see Supplement). We then implemented a transfer matrix method to calculate reflection and transmission which were used to derive absorption\cite{Sipe_1987}. Emissivity was then calculated via Kirchhoff's law of thermal radiation, equating emissivity to absorptivity, and the overall thermal emission was calculated by integrating over the Planck distribution in wavelength, angle, and polarization \cite{Baranov_Xiao_Nechepurenko_Krasnok_Alu_Kats_2019}.

Figure \ref{fig:intensities}(a-c) displays $I_A(T)$ and $I_E(T)$ for the total structure as well as the constituent materials. At all temperatures, SiO$_2$ dominates thermal emission and roughly follows a $T^4$ power law. When the metasurface is made with Hero SiO$_2$, an incident intensity of $I_0=1$\,GW$\cdot$m$^{-2}$ yields two equilibrium temperatures of 177 K and 516 K with the former being stable and the latter unstable. Above 516 K, thermal runaway occurs as the sail absorbs more energy than it can emit. Increasing the incident intensity to $I_0=10$\,GW$\cdot$m$^{-2}$, we found no equilibrium temperature exists in Hero SiO$_2$, and thermal runaway occurs at all temperatures. For metasurfaces using Typical SiO$_2$ and $I_0=1$\,GW$\cdot$m$^{-2}$, the SiO$_2$ absorption dominates over the TPA in Si at low temperatures.  We found that the stable and unstable equilibrium temperatures are very close at 413\,K and 475\,K. This means a small fluctuation in incident intensity could put the sail into the thermal runaway region and cause it to melt. Figure \ref{fig:intensities}d displays the difference in emissive intensity $I_E$ and absorptive intensity $I_A$ as a function of both $T$ and $I_0$ for the case of Hero SiO$_2$. There is a large region where this value is positive indicating the sail will cool to the stable equilibrium temperature, provided the initial temperature is below the unstable equilibrium temperature. We found that no stable equilibrium temperature exists for $I_0>$5.4\,GW$\cdot$m$^{-2}$ implying that the temperature will increase until it sublimates, melts or otherwise fails.

\vspace{-10pt}
\section*{\label{sec:disc}Discussion} 
\vspace{-10pt}

The thermal runaway of the sail that we predict at high temperatures is caused by the increasing free-carrier and bandgap absorption of Si with temperature.  These absorption mechanisms strengthen with increasing temperature because the band gap of Si shrinks and the population of thermally excited carriers increases.  At low temperatures, if the low-loss Hero SiO$_2$ is used, TPA in Si dominates the absorption of the sail.  Since TPA increases linearly with incident power, the absorbed power of the sail increases with the square of the incident power, i.e. $I_A \propto I_0^2$. Strong TPA at low temperatures is sufficient to raise the temperature of the sail to the point where the increasing free-carrier and bandgap absorption cause a thermal runaway.  In principle, it would be possible to avoid TPA by choosing a laser wavelength longer than 2\,$\mu$m, but multi-phonon absorption in the SiO$_2$ will increase, as will free carrier absorption in the Si.  Detailed modeling beyond 2\,$\mu$m, however, is inhibited by a lack of absorption data in Hero-quality SiO$_2$ and Si at those wavelengths. Below 2\,$\mu$m, all laser sail designs containing Si must contend with TPA. Moreover, while operating at low laser powers can decrease TPA and allow an equilibrium temperature to exist, there is a danger that a small localized transient temperature fluctuation in the sail could lead the entire sail into thermal runaway.

The total emission of the sail could be increased to counteract the effects of TPA. Laser cooling of rare-earth ions \cite{Jin_Guo_Orenstein_Fan_2021} or nanoparticle laminate films \cite{Wray_Su_Atwater_2020} has been proposed.  Additionally, the emission spectrum of coupled resonators has been shown to be highly tunable,\cite{Morsy_OE_2021} and such techniques could be applied to increase the emission of a sail incorporating SiO$_2$. 

The maximum laser intensities set by the thermal limits we calculate here also place limits on the maximum achievable acceleration of the sail. Laser sail designs in other works assumed $I_0\approx$25\,GW$\cdot$m$^{-2}$\cite{ilic_went_atwater_2018,Salary_Mosallaei_2020}. In this work, we find that a Si/SiO$_2$ sail will be limited to $I_0<$5.4\,GW$\cdot$m$^{-2}$ using the best materials available. A commonly used figure of merit for sail performance is the smallest possible acceleration distance $D$ for a given incident intensity and target speed\cite{Jin_Li_Orenstein_Fan_2020}.  We used the reflection spectrum in Fig. \ref{fig:absem}c to calculate $D$ for the sail studied here and found that at $I_0=25$\,GW$\cdot$m$^{-2}$, $D=26$\,Gm, but that reducing I$_0$ to the thermally limited value of $I_0=5.4$\,GW$\cdot$m$^{-2}$ yields $D=120$\,Gm, a penalty of a factor of 5. We note that while our sail was not optimized for acceleration, $D$ scales as $1/I_0$, so the penalty will be roughly the same factor for an optimized sail.

 It is possible that light sails constructed of a different material would not exhibit the thermal runaway behavior of Si sails. In particular, silcon nitride has been shown to be a promising alternative \cite{Myilswamy_Krishnan_Povinelli_2020}. A thermal stability analysis of a sail made from SiN$_x$ would be of great interest. However, we are unaware of any measurements of the temperature-dependent absorption of SiN$_x$ at 1.55$\mu$m from 300--800\,K, making a determination of the thermal stability beyond the scope of this work.  
 
\vspace{-10pt}
\section*{\label{sec:disc}Conclusion} 
\vspace{-10pt}

In this work, we demonstrated that a Si/SiO$_2$ metasurface floating in vacuum and exposed to a 1.55\,$\mu$m laser with an intensity $>$\,5.4\,GW$\cdot$m$^{-2}$ will melt, regardless of starting temperature. Equilibrium temperatures exist at lower incident laser intensities; however, a thermal runaway process will melt the sail if the sail temperature reaches the 400--500\,K range. The use of high quality SiO$_2$ with low absorption will increase the thermal stability of the sail to a point, but ultimately the absorption will be dominated by two-photon absorption in Si. Thus, the potential for thermal runaway must be taken into account when designing a laser sail that incorporates a temperature-dependent absorptivity such as Si.

\section*{Acknowledgements}

This work is supported by the Breakthrough Initiatives, a division of the Breakthrough Prize Foundation, and by the Gordon and Betty Moore Foundation through a Moore Inventors Fellowship.

\section*{Supporting Information}
The supporting information document details the fitting procedure for the multiphonon modes and the interpolation between the different literature models for the Si composite absorption model presented here. In addition, the analysis performed in the main text is repeated for an unpatterned Si/SiO$_2$ heterostructure. Finally, we provide a validation for our use of an effective medium approximation when calculating the emissivity of the metasurface.

\bibliography{starref}

\end{document}